\documentclass[twocolumn,showpacs,superscriptaddress,twoside,prl]{revtex4}
\usepackage{amsmath,amssymb,url,graphicx,epstopdf,bm}
\usepackage{braket}
\usepackage{epsfig,amsmath,amssymb,amsfonts, latexsym, dsfont, wasysym,multirow,mathrsfs,xcolor,color}
\usepackage[colorlinks=true,citecolor=blue,urlcolor=blue,linkcolor=blue]{hyperref}
\usepackage[caption=false]{subfig}

\begin{document}

\title{High-Fidelity Single-Shot Toffoli Gate via Quantum Control}
\author{Ehsan Zahedinejad}\email{ezahedin@ucalgary.ca}
\affiliation{Institute for Quantum Science and Technology, University of Calgary, Alberta, Canada T2N 1N4}
\author{Joydip Ghosh}\email{ghoshj@ucalgary.ca}
\affiliation{Institute for Quantum Science and Technology, University of Calgary, Alberta, Canada T2N 1N4}
\author{Barry C. Sanders}\email{sandersb@ucalgary.ca}
\affiliation{Institute for Quantum Science and Technology, University of Calgary, Alberta, Canada T2N 1N4}
\affiliation{Program in Quantum Information Science, 	Canadian Institute for Advanced Research, Toronto, Ontario M5G 1Z8, Canada}
\affiliation{Hefei National Laboratory for Physical Sciences at the Microscale and Department of Modern Physics, University of Science and Technology of China, Anhui 230026, China}
\affiliation{Shanghai Branch, CAS Center for Excellence and Synergetic Innovation Center in Quantum Information and Quantum Physics, University of Science and Technology of China, Shanghai 201315, China}

\begin{abstract}
A single-shot Toffoli, or controlled-controlled-NOT,
gate is desirable for classical and quantum information processing. The Toffoli gate alone is universal for reversible computing and, accompanied by the Hadamard gate, forms a universal gate set for quantum computing. The Toffoli gate is also a key ingredient for (non-topological) quantum error correction. Currently Toffoli gates are achieved by decomposing into sequentially implemented single- and two-qubit gates, which requires much longer times and yields lower overall fidelities compared to a single-shot implementation. We develop a quantum-control procedure to construct a single-shot Toffoli gate for three nearest-neighbor-coupled superconducting transmon systems such that the fidelity is 99.9\% and is as fast as an entangling two-qubit gate under the same realistic conditions. The gate is achieved by a non-greedy quantum control procedure using our enhanced version of the Differential Evolution algorithm.
\end{abstract}

\date{\today}
\maketitle

Scalable quantum computing~\cite{Got98,NC05} 
requires a set of high-fidelity universal quantum gates with which to construct the circuit~\cite{NC05,BKM+14,CGJ+13}.
Experimental progress towards a high-fidelity universal set of gates comprising
single- and two-qubit operations has been impressive, exceeding 99.9\% for single-qubit gates
and 99\% for an entangling two-qubit gate~\cite{BKM+14},
but an outstanding problem is that (non-topological) quantum error correcting codes
require a gate acting on at least three qubits~\cite{CPM+98,RDN+12},
with the Toffoli gate~\cite{MKH+09,FSB+12,SFW+12} being optimal.
The Toffoli gate is also a key component for reversible arithmetic operations, 
such as the modular exponentiation, which is a necessary step in Shor's factoring algorithm~\cite{VSB+01}.

The quantum Toffoli gate is to effect a three-qubit controlled-controlled-NOT (CCNOT) gate,
which means that the third qubit is flipped
only if the first two qubits are in the~$|1\rangle$ state and not flipped otherwise.
Thus far Toffoli gates are achieved by decomposing into single- and two-qubit gates
with resultant fidelities limited to 81\% in a post-selected photonic circuit~\cite{LBA+09},
71\% in an ion-trap system~\cite{MKH+09},
68.5\% in a three-qubit circuit QED system~\cite{FSB+12},
and 78\% in a four-qubit circuit QED system~\cite{RDN+12}.
We here introduce a non-greedy quantum-control approach for directly constructing Toffoli gates
based on an enhanced version of the Differential Evolution (DE) algorithm~\cite{LCP+13,ZSS+14}.
We show that our scheme applied to the three nearest-neighbor-coupled superconducting transmon systems should produce a Toffoli gate operating with 99.9\% fidelity and operating as fast as an entangling two-qubit gate under the same conditions. As our quantum-control-based approach~\cite{PDR88} to realizing the Toffoli gate does not resort to decomposition, a fast Toffoli gate enables error-correction with high fidelity under this scheme.
An additional valuable benefit of realizing CCNOT directly is that
the Hadamard ($H$) and CCNOT together make a universal gate set~\cite{Shi03}
with significant advantages over the oft-studied $H$,~$\pi/8$ gate and CNOT universal set~\cite{NC05}.

Superconducting circuits offer a promising medium for realizing a high-fidelity CCNOT gate
based on quantum control 
of three nearest-neighbor-coupled superconducting artificial atoms~\cite{BKM+14}.
Our approach is to vary the energy levels
for each of three individual superconducting atoms 
using time-dependent control electronics,
which conveniently do not require additional microwave control~\cite{GGZ+13}.
A similar strategy has recently been employed successfully to design two-qubit controlled-Z gates, 
for which optimal pulses are found via a greedy algorithm~\cite{EW14}.
We, however, have observed that existing optimization algorithms (including greedy algorithms) are insufficient to generate an optimal pulse for high-fidelity Toffoli gates, and, therefore, developed a non-greedy optimization scheme, 
referred to here as Subspace-Selective Self-Adaptive DE or SuSSADE.

We consider a linear chain of three nearest-neighbor-coupled superconducting 
artificial atoms, realized as transmons~\cite{BKM+14}
with distinct locations labeled $k=1,2,3$.
The transmons have non-degenerate discrete energy levels,
labeled~$\{|j\rangle_k\}$, with $j=0$ for the ground state.
The energies are anharmonically spaced, with this spacing 
allowed to be dependent on the specific transmon. 
Whereas superconducting atoms contain many energy levels, we truncate all energy levels 
for $j>3$ for each transmon as a CCNOT operates on at most three excitations.

Although the Toffoli gate acts on three qubits as per definition,
our quantum-control procedure operates on the first four levels of each transmon.
The Hamiltonian that generates Toffoli acts on the $4^3$-dimensional Hilbert space~$\mathscr{H}_4^{\otimes 3}$
with energy basis $\{|j\in\{0,1,2,3\}\rangle^{\otimes 3}\}$.
We follow the standard practice of specifying transmon energy levels instead as frequencies
with these atomic frequencies shifted by frequency of a rotating-frame basis transformation:
the shifted frequency of the~$k^\text{th}$ transmon is~$\Delta_k$,
and the anharmonicity of the~$j^\text{th}$ level of the~$k^\text{th}$ transmon is~$\eta_{jk}$.
Therefore,
the energy of the~$k^\text{th}$ transmon's~$j^\text{th}$ level at time~$t$
is $h(j\Delta_k(t)-\eta_{jk}$).

Nearest-neighbor transmons couple via an~$XY$ interaction
with coupling strength
between the~$k^\text{th}$ and~$(k+1)^\text{th}$ transmons
denoted by~$g_k$. 
The three-transmon Hamiltonian is thus~\cite{GGZ+13}
\begin{align}
\label{eq:QCMHamiltonian}
\frac{\hat{H}(t)}{h} =& \sum_{k=1}^3\sum_{j=0}^3\left(j\Delta_k(t)-\eta_{jk}\right)\ket{j}_k\bra{j}_k \nonumber \\
			&+\sum_{k=1}^2\frac{g_k}{2}\left(X_kX_{k+1}+Y_kY_{k+1}\right),
\end{align}

for coupling operators
\begin{align}
\label{eq:xdef}
X_k =& \sum\limits_{j=1}^3\sqrt{j}\ket{j-1}_k\bra{j}_k + \text{hc}, \nonumber \\
Y_k =& -\sum\limits_{j=1}^3\sqrt{-j}\ket{j-1}_k\bra{j}_k + \text{hc},
\end{align}
which are higher-dimensional generalizations of Pauli spin matrices~\cite{BdGS02,GGZ+13},
and~hc denotes Hermitian conjugate.

Here we employ Hamiltonian evolution to realize the CCZ gate,
which effects $\alpha|0\rangle+\beta|1\rangle\mapsto\alpha|0\rangle-\beta|1\rangle$
on the third qubit
only if the first two qubits are~$|1\rangle$.
The CCNOT and CCZ operations are equivalent under the local transformation
CCNOT$=\left[\mathds{1}{\otimes}\mathds{1}\otimes H\right]
	\text{CCZ}\left[\mathds{1}{\otimes}\mathds{1}\otimes H\right]$ (similar to the equivalence between two-qubit CNOT and CZ gates), with~$H$ straightforward to implement as a fast single-qubit operation~\cite{KBC+14,MGR+09}.
The CCZ gate is achieved by varying~$\Delta_k$ of each superconducting atom
over duration~$\Theta$ with resultant Hamiltonian-generated time-ordered ($\mathcal T$)
evolution operator
\begin{equation}
\label{eq:UTheta}
	U(\Theta)=\mathcal{T}\exp\left(-\text{i}\int\limits_0^{\Theta}\hat{H}(\tau)\text{d}\tau\right).
\end{equation}

Whereas our approach enables generating any desirable pulse shape for $\Delta_k$, here we consider two types of pulses: piecewise-constant and piecewise-error-function.
These time-dependent control pulses are constrained within the frequency-range of a superconducting transmon system.
We employ the less computationally-expensive piecewise-constant function to demonstrate the existence of an optimal pulse for high-fidelity Toffoli gate.
However, the control electronics for superconducting systems is only capable of generating smooth pulses, which motivated us to consider a realistic case for which the control parameters are connected together via smooth error functions~\cite{GGZ+13,GKM12}. We show that the gate fidelity does not depend on the pulse shape, and only depends on the number of control parameters. Therefore, without any loss of generality, we choose the less computationally expensive piecewise-constant control function to analyze the fidelity of the designed gate against other parameters.

The Hamiltonian evolution~(\ref{eq:UTheta}) describes the unitary dynamics of the system in the absence of decoherence.
Decoherence is incorporated by treating each atom as a damped harmonic oscillator
characterised by amplitude and scattering induced phase-damping rates for each oscillator.
The corresponding timescales are relaxation time~$T_1$ and dephasing time~$T_2$,
analogous to the rates employed for two-level systems~\cite{NC05,LOM+04}.
We here assume $T:=T_\text{1}\equiv T_\text{2}$,
which is valid for frequency-tunable transmons~\cite{GGZ+13}.
These decohering processes modify the unitary evolution~(\ref{eq:UTheta})
to a completely-positive trace-preserving map~$\mathcal{E}(\Theta)$,
which is decomposable into an operator sum
as discussed in the Supplementary Material~\cite{RefSupplementary}.

A high-fidelity quantum gate is usually designed
by determining an optimal control pulse for each frequency~$\Delta_k$
neglecting open-system effects such as decoherence.
Performance is assessed for the unitary evolution~(\ref{eq:UTheta})
projected to the computational subspace:
$U_\mathscr{P}(\Theta):=\mathscr{P}U(\Theta)\mathscr{P}$.
The standard figure of merit for performance of $U_\mathscr{P}(\Theta)$
is the ``intrinsic fidelity'' (fidelity neglecting decoherence)
with respect to the ideal gate, in this case CCZ,
so the intrinsic fidelity is~\cite{ZSS+14}
$\mathcal{F}
		=\frac{1}{8} \left|\text{Tr}\left(\text{CCZ}^{\dagger}\;U_\mathscr{P}(\Theta)\right)\right|$
with $\mathcal{F}=1$ if $U_\mathscr{P}(\Theta)=$ CCZ
and $0\leq\mathcal{F}<1$ otherwise.
After determining control pulses that maximize~$\mathcal F$,
decoherence is then incorporated into the calculation
to assess the performance under open-system conditions~\cite{GGZ+13}.

In the presence of decoherence,
the efficacy of the non-unitary evolution compared to the target gate is quantified
by the average state fidelity~$\bar{\mathcal{F}}$,
which is calculated as follows.
For $\{\ket{\psi_k}\in\mathscr{H}_2^{\otimes 3}\}$
the set of three-transmon computational basis states,
the non-unitary extension of the unitary evolution~(\ref{eq:UTheta})
transforms a pure computational basis state $|\psi_k\rangle\langle\psi_k|$
into a mixed state $\rho^\text{final}_k$. 
As each basis state $\ket{\psi_k}\bra{\psi_k}$ remains invariant under an ideal CCZ gate,
average state fidelity
$\bar{\mathcal F}=\frac{1}{8}\sum_k\sqrt{\left|\bra{\psi_k}\rho^\text{final}_k\ket{\psi_k}\right|}$
quantifies the efficacy of a quantum gate in the presence of intrinsic
as well as decoherence-induced noise for a given optimal pulse. 
Whereas $\bar{\mathcal F} \approx 99.9\%$ is considered to be a threshold
for topological (surface-code) fault-tolerance for single- and two-qubit gates~\cite{BKM+14},
our approach achieves this fidelity even for
the three-qubit CCZ gate subject to realistic constraints of the control pulses.
In this work, unless otherwise stated, the average state fidelity is referred to as fidelity.

The strategy for controlling the evolution~(\ref{eq:UTheta})
is to vary the frequencies so that energy levels approach each other but then avoid degeneracies,
known as avoided level crossings.
These avoided crossings mix energy-level populations and dynamical phases together.
This avoided-crossing effect enables shaping the evolution toward
the final time-evolution operator objective,
which is obtained by maximizing~$\mathcal F$.

Optimal pulse shapes for each $\Delta_k$ 
are obtained by discretizing the time duration~$\Theta$ into~$N$ constant intervals of duration~$\Delta\tau:=\Theta/N$,
and the control-problem parameter space
is spanned by the set of variables
$\{\Delta_k(\ell\Delta{\tau});\ell=1,\dots,N\}$ for each~$k$.
These control-points are then connected via step functions or error functions to construct the pulse shapes as we described earlier.
The CCZ optimization problem is non-convex with a $3N$-dimensional parameter space
corresponding to~$N$ parameters for each of the three frequencies~$\Delta_\text{1,2,3}$.
For a fixed $\Delta\tau$, therefore, the dimension of the parameter space increases linearly with the total time duration $\Theta$, which influences which optimization methods work and which do not.

We devise a quantum-control procedure that designs an optimal pulse for a Toffoli gate, which operates
as fast as a two-qubit gate~\cite{GGZ+13} with a target intrinsic fidelity of 0.9999.
We first use the existing optimization algorithms, namely quasi-Newton approach,
which employs the Broyden-Fletcher-Goldfarb-Shanno
(BFGS) approximation of the
Hessian~\cite{Bro70,Fle70,Fle13,Gol70,Sha70},
simplex methods~\cite{ON75},
several versions of particle swarm optimization~\cite{Tre03,CK02},
and differential evolution (DE) algorithms~\cite{MVC06,ZSS+14}; however, they all failed to
reach the target intrinsic fidelity under the time-constraint of the fast Toffoli gate. Therefore, we construct a new optimization algorithm here
to realize a Toffoli gate that reaches our target.

Of these optimization approaches,
DE yielded the best fidelity but failed to reach the target due to the well known problems 
of searching high-dimensional parameter spaces~\cite{ZKX07}.
This drawback motivated us to enhance
DE for such high-dimensional problems
by instead breeding over randomly selected low-dimensional subspaces,
hence our name
Subspace-Selective Self-Adaptive DE (SuSSADE) algorithm
(see Supplementary Material~\cite{RefSupplementary}).
One of our objectives is to demonstrate (numerically) the capability of SuSSADE with respect to finding a solution equally successfully regardless of parameter-space dimension within the regime that is relevant for current superconducting experiments. 

To understand our enhancement,
we first briefly review standard DE~\cite{MVC06}.
DE cooperatively evolves a collection of trial solutions,
called chromosomes,
towards an optimal solution.
Chromosomes are labeled by their location in the parameter space,
and optimization is thus a search for the best chromosome in this space.
Evolution from one generation (i.e., chromosomes for one iteration step)
to the next is achieved by breeding each chromosome with 
three other randomly chosen chromosomes from the same generation.
Breeding yields a single daughter chromosome,
and only the fittest of the original and daughter chromosome survives.
This breeding-and-survival procedure continues 
until either a chromosome reaches the requisite~$\mathcal F$
or the number of generations reaches a specified upper bound.

Whereas standard DE breeds chromosomes randomly selected from the entire space,
our SuSSADE algorithm is much faster due to breeding being restricted some of the time to a subset of chromosomes drawn from a low-dimensional subspace,
i.e., some fixed parameters and some variable parameters.
Our algorithm randomly switches breeding between the subspace and the whole space
according to the value of an input switch parameter~$S\in[0,1]$
such that a uniformly distributed random number $r_j\in[0,1]$ at generation~$j$
restricts breeding to the subspace if $r_j<S$ and breeds in the whole space otherwise.

In the extreme case of restricting to one-dimensional subspaces,
chromosomes can breed only if all but one of the parameters are the same.
We refer to this one-dimensional extreme case as 1DSuSSADE.
Henceforth we use only 1DSuSSADE as it works well with $S=0.14$ for designing the Toffoli gate.

Here we present two types of pulses that achieve the target intrinsic fidelity, and we
explore how the performance of piecewise constant pulses vary with respect to the
total gate time, coupling strength and decoherence-induced noise.
The success of our quantum control procedure corresponds to a target intrinsic fidelity of $0.9999$ and a timescale comparable to a two-qubit gate~\cite{GGZ+13}.

\begin{figure}
\centering
	\includegraphics[width=0.9\linewidth]{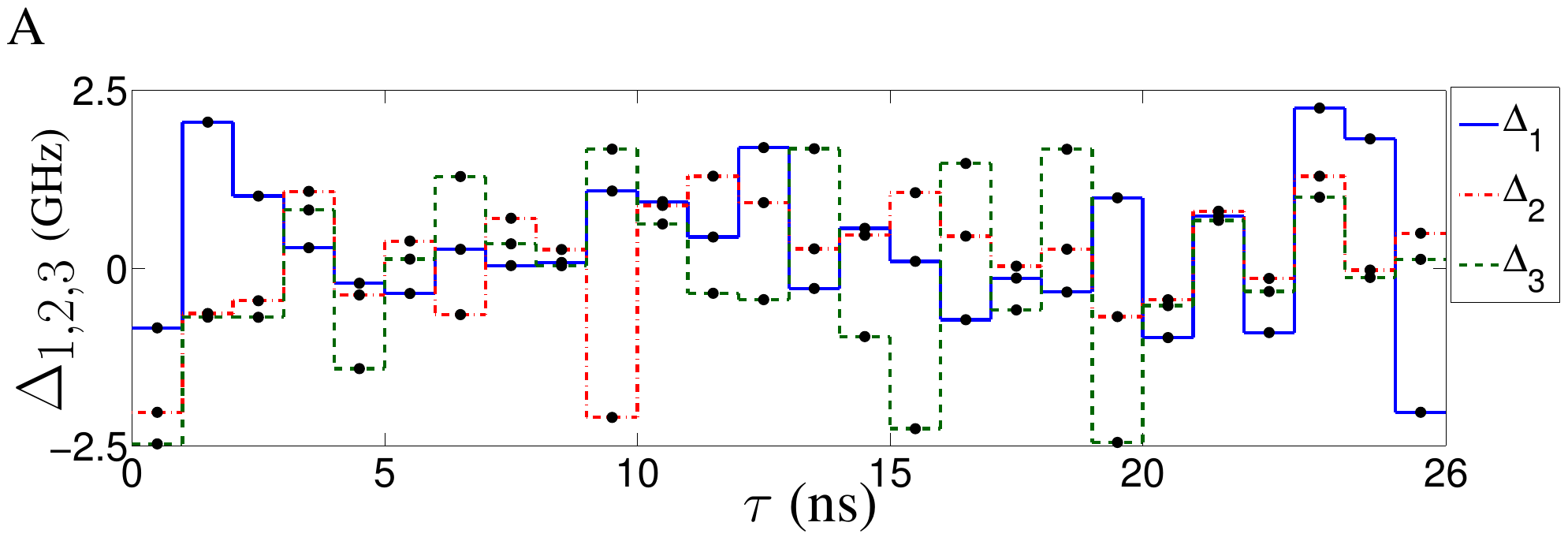} \\
	\includegraphics[width=0.9\linewidth]{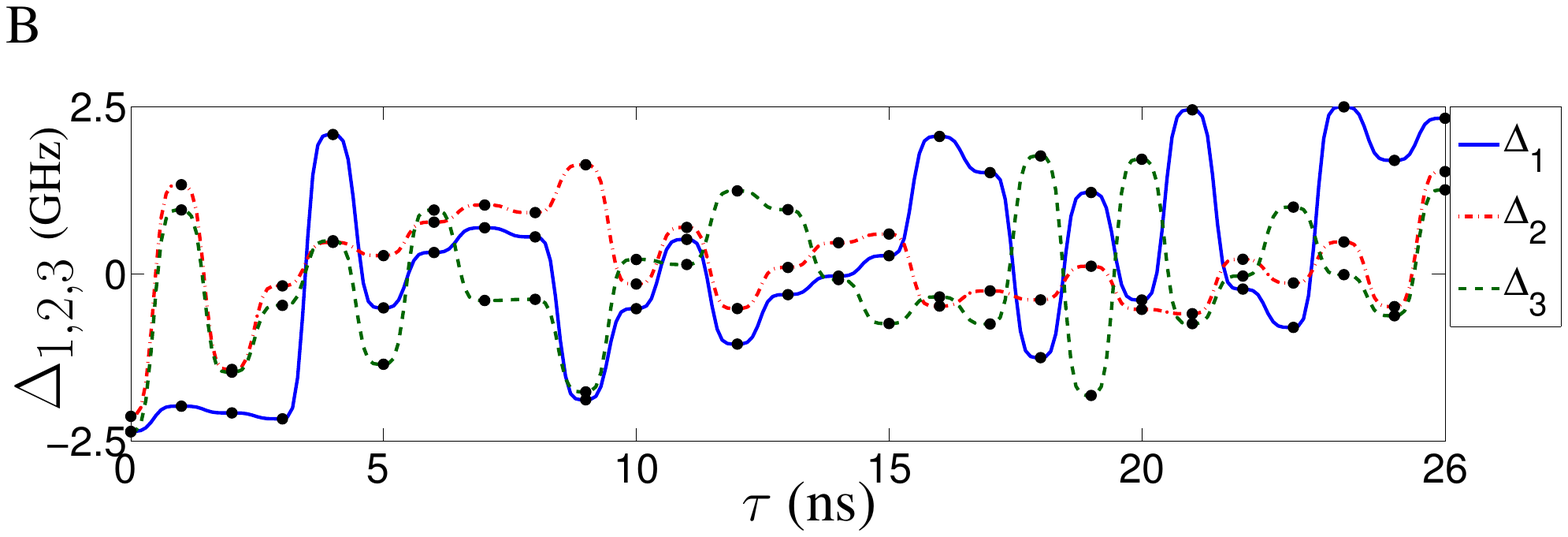}
\caption{(color online)
	Optimal pulse shapes for the Toffoli gate given as frequency detunings~$\Delta_i$
	(for the $i=1,2,3$) of the superconducting atoms, corresponding to A) piecewise-constant and B) error-function-based pulse profiles, as a function of time~$\tau$ with constant step-size time interval $\Delta{\tau}=1$ ns
	and with ${\mathcal F}= 0.9999$ and $g=30$~MHz. The black dots on both plots show the control parameters used to optimize the shape of the pulses.}
\label{fig:pulseShape}
\end{figure}

\begin{figure}
\centering
	\includegraphics[width=.47\linewidth]{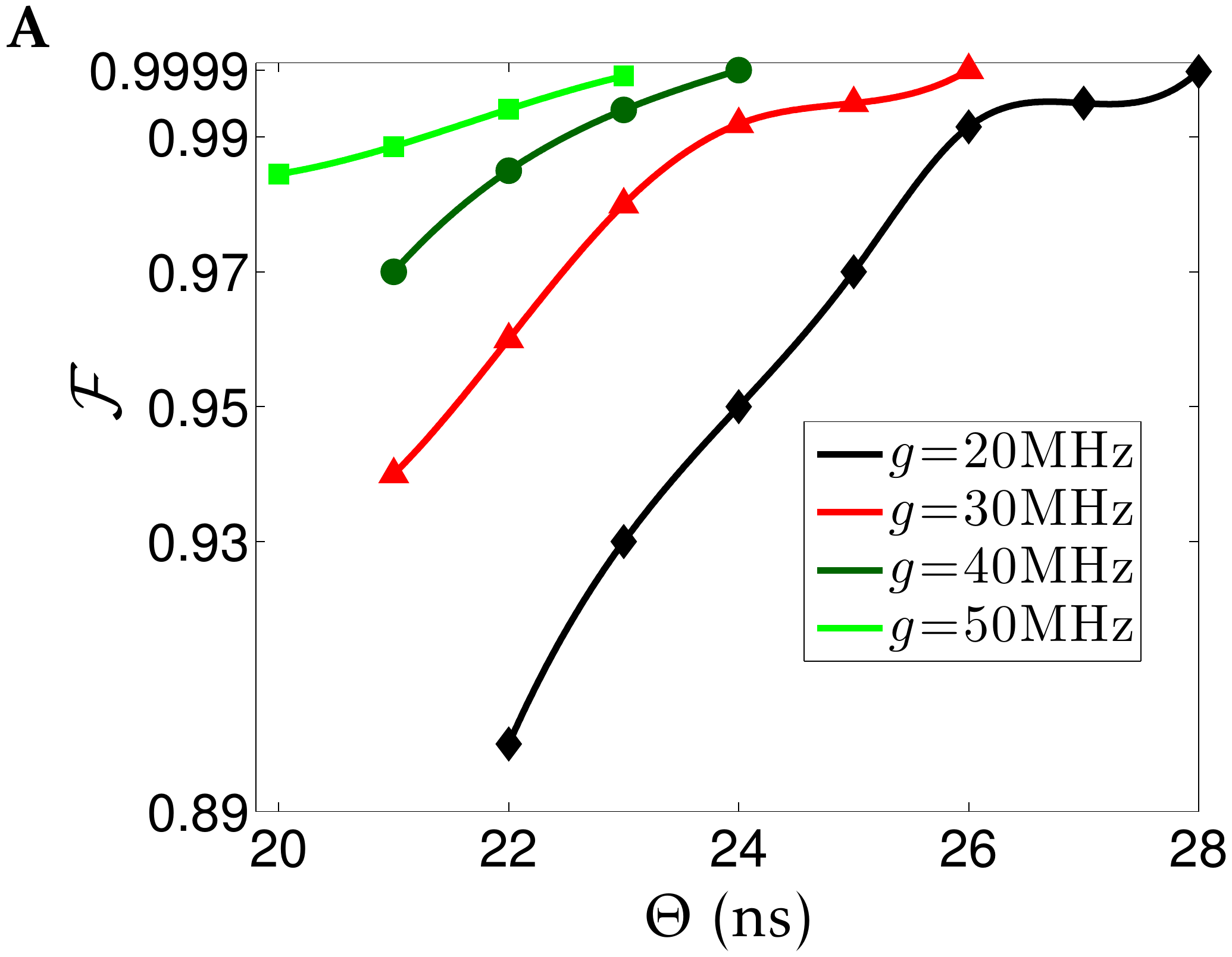}
	\includegraphics[width=.47\linewidth]{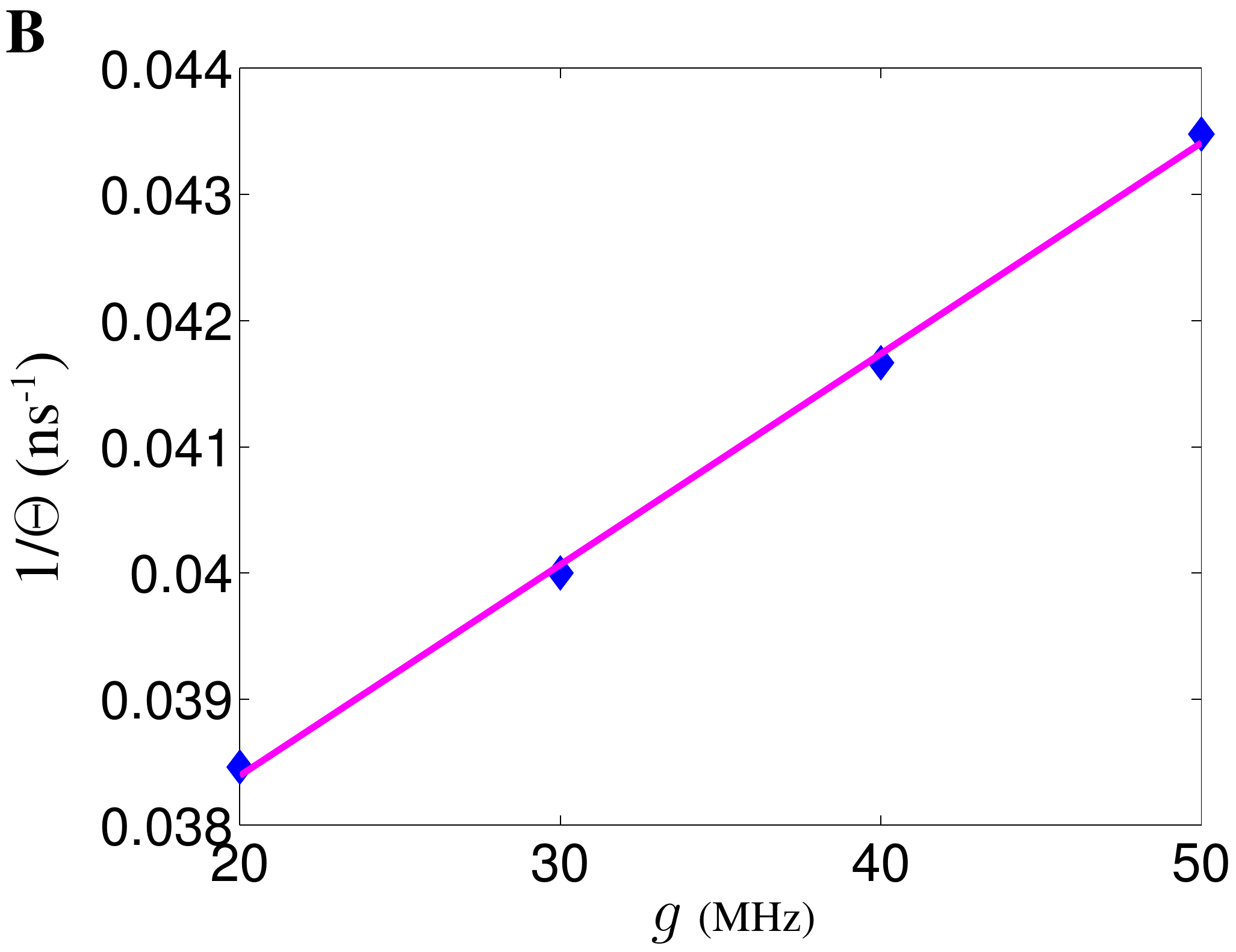}
\caption{(color online)
	A) Intrinsic fidelity~$\mathcal{F}$ vs.~total gate time~$\Theta$
	for various coupling strengths~$g$
	and
	B)~$1/\Theta$ vs~$g$ for ${\mathcal F}=0.999$.
	The $\Diamond$, $\triangle$,  $\circ$, and $\square$ denote the actual numerical computations using 1DSuSSADE, and solid lines depict cubic-fit curves.
	}
\label{fig:fidelityCurves}
\end{figure}

\begin{figure}
\centering
	\includegraphics[width=.7\columnwidth]{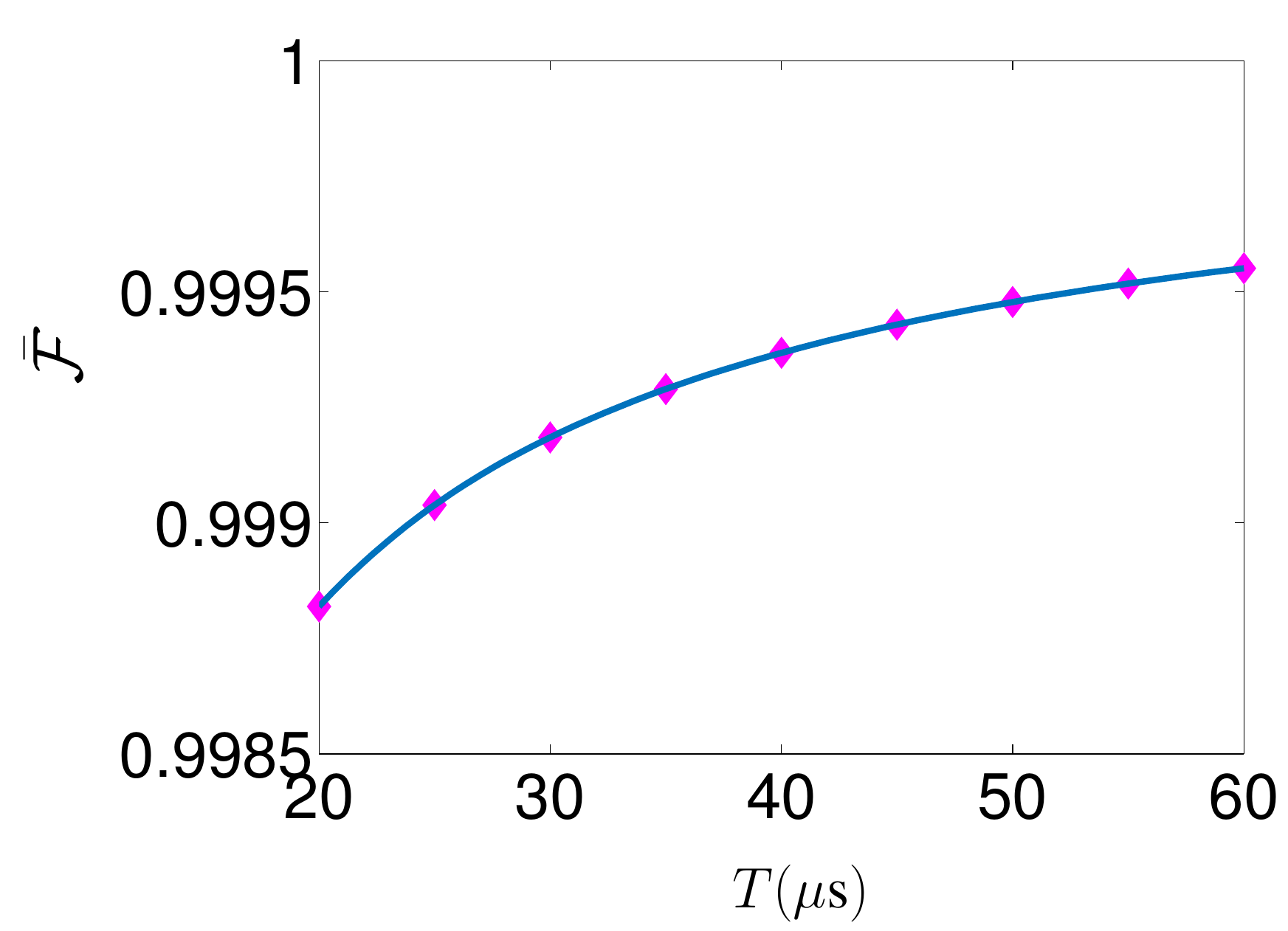}
\caption{(color online)
	The fidelity $\bar{\mathcal F}$
	is plotted against the coherence times $T$. We assume $T=T_\text{1}=T_\text{2}$, with $T_1$ and $T_2$ the relaxation time and dephasing time of each transmon respectively. This assumption is valid for tunable transmons. Each $\Diamond$ denotes an actual numerical result obtained from the decoherence calculation. The solid line depicts the cubic-fit curve.
	}
\label{fig:FvsT}
\end{figure}

Figure~\ref{fig:pulseShape} shows both piecewise-constant as well as piecewise-error-function pulses that achieve the target intrinsic fidelity obtained by optimizing all the parameters within the experimental constraints. 
The CCZ gate corresponding to Figure~\ref{fig:pulseShape} requires a total gate time of~$26$~ns given a coupling strength of $g=30$~MHz~\cite{EW14,GGZ+13}. Comparing the intrinsic fidelities of Figure~\ref{fig:pulseShape}(a) and Figure~\ref{fig:pulseShape}(b) shows that the target fidelity does not depend on the shape of the pulse; rather it depends on the number of control parameters. In what follows, therefore, we consider only the piecewise-constant pulse shapes as these are computationally less expensive to handle and also do not compromise the generality of our results.
We show in Figure~\ref{fig:fidelityCurves} how the (maximized) intrinsic fidelity changes when the parameters $g$ and $\Theta$ are varied within a range commensurate with currently available superconducting circuits~\cite{BKM+14}. Figure~\ref{fig:fidelityCurves}(a) gives the intrinsic fidelity as a function of total gate time for various coupling strengths~$g$.
Figure~\ref{fig:fidelityCurves}(b) shows that the total gate time changes linearly with the coupling~$g$, in order to achieve a given fidelity.
Finally, we consider the effect of decoherence on the approximate CCZ gate obtained 
by optimal pulses shown in Figure~\ref{fig:pulseShape}(a), and compute the fidelity $\bar{\mathcal F}$.
Amplitude-damping and phase-damping rates ($T_1^{-1}$ and $T_2^{-1}$)
are treated as the dominant forms of decoherence.
For fast gates with $\Theta \ll T$, but with decoherence more significant than intrinsic errors, an order-of-magnitude estimate yields $1-\bar{\mathcal F} \sim \Theta/T$~\cite{GGZ+13},
which is consistent with the numerically evaluated plot of~$\bar{\mathcal F}$ vs~$T$
in Figure~\ref{fig:FvsT}.
We have employed our quantum control procedure to determine the optimal pulse (Figure~\ref{fig:pulseShape}) for a high-fidelity single-shot three-qubit Toffoli gate.
We computed a smooth pulse (Figure~\ref{fig:pulseShape}(b)), for which the control parameters are separated by $1$~ns and connected via error functions, thereby ensuring that the pulse is compatible with the power and bandwidth specifications of standard control electronics.
Applying our approach to the three nearest-neighbor-coupled superconducting transmon systems
produces a fast and high-fidelity Toffoli gate in $26$ ns, which matches the timescale for the two-qubit avoided-crossing-based CZ gate.

The longer total gate time, with a fixed $\Delta\tau$,
generates a higher-dimensional parameter space for the optimization algorithm.
The monotonically increasing optimized intrinsic fidelity in Figure~\ref{fig:fidelityCurves}(a) with increasing total time thus demonstrates the capability of our algorithm for 
a parameter regime relevant to superconducting experiments,
for which alternative algorithms fail.
The linear relationship between $1/\Theta$ and $g$
demonstrates that faster gate speed requires higher coupling.
This relation is a characteristic signature for avoided-crossing-based gates (as also obtained for avoided-crossing-based two-qubit gates~\cite{GGZ+13}) assuming the corresponding optimization algorithm is capable of finding the optimal solution regardless of the parameter space dimension.

The effect of decoherence on the performance of the optimal Toffoli gate has been explored (Figure~\ref{fig:FvsT}), and we interpret the corresponding result as follows: when $T \gg \Theta$, the effect of thermal noise becomes less dominant, and the fidelity of a quantum gate is almost entirely contributed by the intrinsic fidelity. Here we have been able to design a fast and optimal pulse for a Toffoli gate using our quantum control approach for which the intrinsic fidelity is so high ($\sim{99.99\%}$), that the fidelity ($\bar{\mathcal F} \sim 99.9\%$) is significantly contributed by the decoherence-induced noise, which is also very small (compared to previous realizations) for the state-of-the-art superconducting atoms with $T\sim\;20-60\;{\mu}$s~\cite{BKM+14}.

In summary, we have devised a powerful quantum control scheme, named SuSSADE, to design a fast and high-fidelity single-shot Toffoli gate for a scalable chain of nearest-neighbor-coupled three-transmon system.
The time required for the Toffoli operation is comparable with the timescale of two-qubit avoided-crossing-based CZ gate, which is the key advantage our quantum-control approach proffers compared to decomposition-based approaches requiring many such two-qubit gates to implement a single CCZ operation. Our three-transmon system serves as a module for all 1D and 2D quantum computing architectures~\cite{GFG12}, and therefore one can realize our scheme in a large-scale multi-qubit architecture, if the undesired couplings are turned off~\cite{CNR+14}. Our approach demonstrates the efficacy of SuSSADE for designing quantum gates as well as yielding the concrete example of a three-qubit gate required for scalable quantum-error-correction. 
 
\begin{acknowledgments}
This research was funded by NSERC, AITF, the University of Calgary's Eyes High Fellowship Program, and China's Thousand Talent Program. We also acknowledge the support from WestGrid (www.westgrid.ca) and Compute Canada Calcul Canada.
We thank Alexandre Blais and Simon Nigg for helpful discussions.
J.G. gratefully acknowledges discussions with Michael Geller.
\end{acknowledgments}

\bibliographystyle{apsrev}

\section{Supplemental Material for ``High-Fidelity Single-Shot Toffoli Gate via Quantum Control''}

\subsection{Subspace-Selective Self-Adaptive Differential Evolution (SuSSADE)}
Differential Evolution (DE) was initially introduced by Storn et al.~\cite{SP97} as a global optimization approach and is currently known to be the most efficient of all the evolutionary algorithms~\cite{VT04}.
In its basic form, DE performs three operations namely:
mutation, crossover and selection.
These three operations cooperatively evolve the $N$-dimensional candidate solutions $C_i$, $i\in\{1,\dots,P\}$ toward their optimal position with $N$ and $P$ being the number of optimization parameters in the search space and population size respectively.

The elements of $C_i$ are piecewise-constant functions, where each element is fixed over the time interval $\Delta \tau=\Theta/N$, with~$\Theta$ being the total time needed to perform the unitary operation. DE mutates the candidate solution as follows:
\begin{equation}
 \label{eq:mu}
 D_i=C_{i_1}+\mu\left(C_{i_2}-C_{i_3}\right)
 \end{equation}
 where $i_1$, $i_2$, $i_3$, $i$ $\in[1,P]$ are integers and mutually distinct.
The quantity $\mu\in[0,2]$ is a uniformly distributed random number, which defines the step size
by which DE explores the search space,
and~$D_i$ is the donor vector resulting from the mutation operation~(\ref{eq:mu}).

The next step is to switch the elements $D_i(j)$ of donor vector and each candidate solution $C_i$ via the crossover operation
 \begin{equation}
\label{eq:cr}
	T_i(j)= \begin{cases}
	D_i(j), &\text{if $\chi<\xi$}\\
	C_i(j), &\text{otherwise}
\end{cases}
\end{equation}
with $\chi\in[0,1]$ is a random number chosen from a uniform distribution and $\xi$ defines the crossover rate.
The final step is the selection
\begin{equation}
\label{eq:sl}
	S'_i= \begin{cases}
	C_i &\text{if $F(C_i)<f(X_i)$}\\
	X_i &\text{otherwise}
\end{cases}
\end{equation}
with~$S'_i$ being the offspring of $C_i$ for the next
generation and $f(X_i)$ being the objective function, which is the intrinsic fidelity ${\mathcal F}$ for our case.

Here the standard version of DE was not able to deliver a fidelity better than 95\% for the Toffoli gate.
Thus, we enhance the standard DE algorithm in order to find the optimal solution for problems with high dimension.
The first step towards this enhancement is to find the optimal value of algorithmic parameters, namely $\mu$ and $\xi$~\cite{MSP+11}. One approach is to try DE with many trial guesses for the algorithmic parameters to find the best set for $\mu$ and $\xi$, leading to the best optimal solution for a specific problem under consideration.
However, this trial-and-error approach becomes intractable (as it is in our case) 
when the computational cost of evaluating the objective function is expensive, and running DE for a long time requires an excessive amount of computational resources.

As an alternative approach, we use the self-adaptive DE~\cite{BZM06},
which self-adaptively updates the value of $\mu$ and $\xi$ at each generation $G$ using
\begin{equation}
\label{rate}
	\mu_{i,G+1}= \begin{cases}
	\mu_l+r_1.\mu_u &\text{if $r_2<\kappa_1$}\\
	\mu_{i,G} &\text{otherwise}\\
	\end{cases}
\end{equation}
and
\begin{equation}
\label{rate}
	\xi_{i,G+1}= \begin{cases}
	r_3 &\text{if $r_4<\kappa_2$}\\
	\xi_{i,G} &\text{otherwise,}\\
	\end{cases}
\end{equation}
thereby producing the new algorithmic parameters in a new parent vector
(candidate solution in each generation). 
The quantities $r_j$, with $j \in {1,2,3,4}$, are uniform random numbers generated from $[0,1]$, and $\kappa_1$ and $\kappa_2$ are the probabilities to adjust the algorithmic parameters. 
The quantities~$\kappa_1$,
$\kappa_2$, $\mu_u$ and $\mu_l$ are assigned to fixed values: $0.1$, $0.1$, $0.1$ and $0.9$ respectively.
Whereas using self-adaptive DE by itself improves the obtained fidelity, the resulted fidelity still remains under the requisite threshold ($\sim 99.9\%$) due to the high-dimensionality problem.

The next step to enhance the standard DE is to combine self-adaptive DE with cooperative coevolution (DECC-II)~\cite{ZKX07}, designed for high-dimensional optimization problems.
Even the combination of self-adaptive DE with DECC-II in their original forms
failed to yield a better fidelity so modified this combination.
Here we first describe the DECC-II followed by our enhanced version of DE.

DECC-II works by decomposing the $K$-dimensional candidate solutions to $m$-dimensional subspaces and optimizes each subspace for $s$ cycles at each generation,
whereas other subspaces remain unchanged. DECC-II uses Self-Adaptive Differential Evolution with Neighborhood Search (NSDE)~\cite{YTY08} to optimize each individual subspace. The number~$s$ of cycles  and the dimension~$m$ of the subspace should be assigned by the user, and selecting the optimal value is a computationally expensive task. We found the results to be largely influenced by the choice of $s$ and $m$, with none of them satisfying our requirement of having a threshold-fidelity Toffoli gate.

Inspired by the DECC-II algorithms, we set $s=1$ and choose the dimension of subspace $m$ randomly from \{1, 2, 3, 4, 5\}. Now, we no longer need to look for the optimal values of $s$ and $m$ for the optimization. However this new approach has an extremely slow convergence, as in each generation only a small part of the candidate solutions are being selected for the optimization. In order to speed up the convergence, our algorithm randomly switches breeding between the subspace and the whole space
according to the value of an input switch parameter~$S\in[0,1]$,
such that a uniformly distributed random number $r_j\in[0,1]$ at generation~$j$
restricts breeding to the subspace, if $r_j<S$, and breeds in the whole space otherwise.
As our algorithm selects an $m$-dimensional subspace at each generation and self-adaptively evolves the algorithmic parameters, we call it Subspace-Selective Self-Adaptive DE (SuSSADE). For our purpose, we observe that choosing $m=1$ suffices,
which signifies that the selected subspace is trivial. We refer to this one-dimensional extreme case as 1DSUSSADE.

\subsection{Phase compensation}
In order to design the optimal pulse for the CCZ gate, we employ another strategy, called
``Phase Compensation'', which is frequently used in designing quantum gates with superconducting atoms~\cite{GKM12,GGZ+13}. Any arbitrary rotation about the $z$-axis is trivial for a superconducting qubit and is usually performed via qubit frequency excursions.
Therefore, we assume that any phase acquired by $\ket{100}$, $\ket{010}$, and $\ket{001}$ states can be nullified by post $z$-rotations.

Whereas an ideal CCZ gate is given by (in tensor product basis as defined in the main text), $\text{CCZ}_\text{ideal}=\text{diag}(1,1,1,1,1,1,1,-1)$. Under the freedom that some additional phases can be compensated with post-rotations about $z$-axis, it is sufficient for our purpose if we define our target  gate as
\begin{align}
\label{eq:targetCCZ}
\text{CCZ}:=\text{diag}&\left(1,e^{i\theta_{3}},e^{i\theta_{2}},e^{i(\theta_{2}+\theta_{3})},e^{i\theta_{1}},e^{i(\theta_{1}+\theta_{3})},e^{i(\theta_{1}+\theta_{2})}, \right. \nonumber \\
 &\quad\left.-e^{i(\theta_{1}+\theta_{2}+\theta_{3})}\right),
\end{align}
with $\theta_{j}$ being any arbitrary phase acquired by the $j^\text{th}$ transmon under the optimal pulse.

\subsection{Decoherence}
In order to estimate the effect of decoherence, we model each superconducting transmon as a damped harmonic oscillator and then compute the time evolution of the density matrix under amplitude and phase damping~\cite{NC05,LOM+04}.
We employ Kraus's operator-sum representation to represent the completely-positive trace-preserving maps for these damping processes. If $\rho(t)$ is the density matrix of the system at any given time $t$, then, under a specific dissipative process,
\begin{equation}
	\rho(t)=\sum_{k=0}^{n}{\mathfrak E}_{k}(t)\rho(0){\mathfrak E}^{\dagger}_{k}(t),
\end{equation}
with~${\mathfrak E}_{k}$'s the Kraus matrices for the given process satisfying $\sum_{k=0}^{n}{\mathfrak E}^{\dagger}_{k}{\mathfrak E}_{k}=\mathds{1}$ at each time instant. 

We now discuss construction of Kraus matrices for amplitude and phase damping of a single transmon. Once we know the Kraus matrices for each transmon, the set of Kraus matrices for the entire three-transmon system can then be constructed as all possible tensor products of those single-transmon Kraus matrices, assuming that decoherence affects each superconducting atom independently.

\emph{Amplitude damping:}
The Kraus matrices for amplitude damping of a single superconducting atom are given by (assuming $4$ energy levels for each atom)\cite{LOM+04}
\begin{equation}
	E_{l}(t)=\sum_{j=l}^{3}\sqrt{j \choose l}\left(e^{-\frac{t}{T_{1}}}\right)^{\frac{j-l}{2}}
		\left(1-e^{-\frac{t}{T_{1}}}\right)^{\frac{l}{2}}\ket{j-l}\bra{j}
\end{equation}
with $l\in\{0,1,2,3\}$ denoting the indices for the Kraus matrices. It is straightforward to verify that $\sum_{l=0}^{3}E_{l}^{\dagger}(t)E_{l}(t)=\mathds{1}$ at all time. 
The exponential amplitude-damping factor $\exp(-t/T_\text{1})$ 
has relaxation time~$T_\text{1}$.

\emph{Phase damping:}
The Kraus matrices for phase damping of a single superconducting atom are given by (assuming $4$ energy levels for each atom)~\cite{LOM+04},
\begin{equation}
	{\mathbb E}_{l}(t)=\sum_{j=0}^{3}\exp\left\{-\frac{j^{2}t}{2T_\text{2}}\right\}\sqrt{\frac{(j^{2}t/T_\text{2})^{l}}{l!}}\ket{j}\bra{j},
\end{equation}
with~$l$ again denoting Kraus-matrix indices.
Here~$T_\text{2}$ denotes the dephasing time.
For phase damping,
$\sum_{l=0}^{\infty}{\mathbb E}_{l}^{\dagger}(t){\mathbb E}_{l}(t)=\mathds{1}$ 
at all times is straightforward to check.
Such a completeness relation cannot be satisfied if we truncate the infinite series, 
thereby constraining the Kraus-matrix index~$l$ to a finite upper bound. 
We must impose an upper limit for $l$ in order to ensure a finite number of Kraus matrices.
Such a truncation is valid as long as our total gate time is much shorter than the dephasing time $T_\text{2}$ (implying $t/T_\text{2} \ll 1$), which is true for our case. In our calculation here, we assume $0 \leq l \leq 3$ for phase damping as well, neglecting all higher order terms for $t/T_\text{2}$.

\subsection{Proposed experimental realization}

In this section we discuss the calibration procedure to suppress the effect of residual noise in an experiment, where a fast and high-fidelity Toffoli gate is demonstrated. In a quantum control procedure, the effect of noise is to distort the external pulses e.g. time-dependent qubit frequencies, which are designed optimally through a control procedure. In an ideal experiment with no noise acting on the quantum system, Fig.~\ref{fig:pulseShape}(a) shows our optimal pulses in the absence of distortion. However this idealized assumption is not valid in a real experimental procedure when noise distort the designed pulses. Therefore a calibration procedure must take the distortions into account for optimally designed square pulses. 

	Up to first order, distortion occurs in an experiment whenever a square pulse is generated and passed through the superconducting control electronics.
	Therefore, by the time that the square pulse reaches the artificial atom,
	the pulse becomes smoothed.
	Mathematically, we model first-order distortion by convolving the square pulse
	with a Gaussian function, which yields an error-function (erf) pulse~\cite{GGZ+13,GKM12}.
	Leading-order distortion has already been taken into account in designing our optimal pulses
	as shown in Fig.~\ref{fig:pulseShape}(b) in the original manuscript.\\
	Calibration is required for higher orders and can be achieved
	through the well-known Closed-Loop Learning Control (ClLC) technique~\cite{RVM+00,JR92}.
	ClLC has been successfully applied to tasks such as 
	discriminating similar molecules~\cite{LTR+02}, ionization~\cite{SRV+04},
	molecular isomerization~\cite{KWB+09},
	and coherent quantum control of two-photon transitions~\cite{MS98}.

\begin{figure}
\centering
	\includegraphics[width=0.7\linewidth]{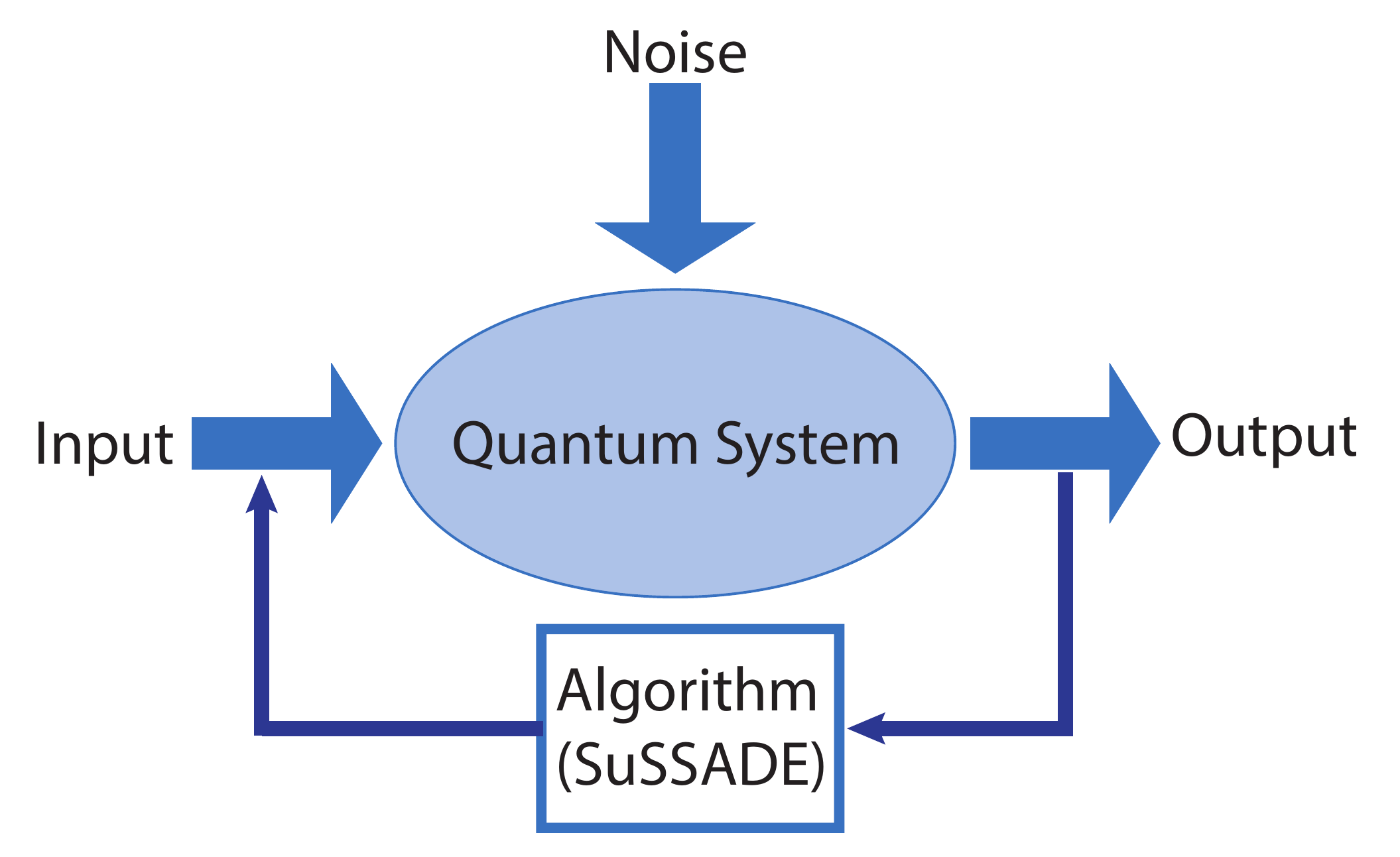}
\caption{(color online)
	A schematic view of the Closed-Loop Learning Control (ClLC) technique which uses SuSSADE as a quantum control scheme. Based on this control procedure, the optimal pulses which are generated using SuSSADE fed into a noisy quantum system. if the output met the target fidelity, the procedure aborts otherwise a new set of control pulses is chosen using SuSSADE, and the iteration continues until the target is met.}
\label{fig:AQF}
\end{figure}

ClLC is an iterative technique for searching the optimal solution in a quantum control
landscape. The idea is to start with a quantum control scheme (SuSSADE for our case) to
solve an ideal, physically realistic model of the quantum system and then test the obtained optimal
solution experimentally. If the experimentally measured objective function
(fidelity in our case) does not satisfy the target, the control pulse should be calibrated again
with the same quantum control algorithm (SuSSADE for our case) in order to obtain a new set of optimal solutions, and this
process continues until the target is met. A schematic view of ClLC is depicted in Fig.~\ref{fig:AQF}.

\end{document}